\documentclass[aps,prb,twocolumn,superscriptaddress,floatfix]{revtex4}

\usepackage{graphicx,graphics}
\usepackage{dcolumn}
\usepackage{amsmath,amssymb,amsfonts}
\usepackage{latexsym,verbatim}
\usepackage{bm}
\usepackage{color}
\usepackage{ulem}
\usepackage{boxedminipage}
\def\be{\begin{equation}}
\def\ee{\end{equation}}
\def\ber{\begin{eqnarray}}
\def\eer{\end{eqnarray}}

\begin{document}
\title{Ultrafast collinear scattering and carrier multiplication in graphene}

\author{D. Brida}
\affiliation{IFN-CNR, Dipartimento di Fisica, Politecnico di Milano, P.za Leonardo da Vinci, 20133 Milano, Italy}
\author{A. Tomadin}
\affiliation{NEST, Istituto Nanoscienze-CNR and Scuola Normale Superiore, I-56126 Pisa, Italy}
\author{C. Manzoni}
\affiliation{IFN-CNR, Dipartimento di Fisica, Politecnico di Milano, P.za Leonardo da Vinci, 20133 Milano, Italy}
\author{Y. J. Kim}
\affiliation{Department of Physics and Astronomy, University of Manchester, Manchester M13 9 PL, UK}
\author{A. Lombardo}
\affiliation{Engineering Department, University of Cambridge, Cambridge, CB3 OFA, UK}
\author{S. Milana}
\affiliation{Engineering Department, University of Cambridge, Cambridge, CB3 OFA, UK}
\author{R. R. Nair}
\affiliation{Department of Physics and Astronomy, University of Manchester, Manchester M13 9 PL, UK}
\author{K. S. Novoselov}
\affiliation{Department of Physics and Astronomy, University of Manchester, Manchester M13 9 PL, UK}
\author{A. C. Ferrari}
\email{acf26@hermes.cam.ac.uk}
\affiliation{Engineering Department, University of Cambridge, Cambridge, CB3 OFA, UK}
\author{G. Cerullo}
\affiliation{IFN-CNR, Dipartimento di Fisica, Politecnico di Milano, P.za Leonardo da Vinci, 20133 Milano, Italy}
\author{M. Polini}
\affiliation{NEST, Istituto Nanoscienze-CNR and Scuola Normale Superiore, I-56126 Pisa, Italy}

\begin{abstract}
Graphene is emerging as a viable alternative to conventional optoelectronic, plasmonic, and nanophotonic materials. The interaction of light with carriers creates an out-of-equilibrium distribution, which relaxes on an ultrafast timescale to a hot Fermi-Dirac distribution, that subsequently cools via phonon emission. Here we combine pump-probe spectroscopy, featuring extreme temporal resolution and broad spectral coverage, with a microscopic theory based on the quantum Boltzmann equation, to investigate electron-electron collisions in graphene during the very early stages of relaxation. We identify the fundamental physical mechanisms controlling the ultrafast dynamics in graphene, in particular the significant role of ultrafast collinear scattering, enabling Auger processes, including charge multiplication, key to improving photovoltage generation and photodetectors.
\end{abstract}

\maketitle
Photonics encompasses the generation, manipulation, transmission, detection and conversion of photons. Applications of photonics are nowadays ubiquitous, affecting all areas of everyday life. Photonic devices, enabled by a continuous stream of novel materials and new technologies, have evolved with a steady increase in functionalities and reduction of device dimensions and fabrication costs. Graphene is emerging as a viable alternative to conventional optoelectronic, plasmonic, and nanophotonic materials. Graphene has decisive advantages\cite{bonaccorso} such as wavelength-independent absorption, tunability via electrostatic doping, large charge-carrier concentrations, low dissipation rates, high mobility, and the ability to confine electromagnetic energy to unprecedented small volumes\cite{koppens}. These unique optoelectronic properties make it an ideal platform for a variety of photonic applications\cite{bonaccorso}, including fast photodetectors\cite{xia,vicar}, transparent electrodes in displays and photovoltaic modules\cite{bonaccorso,bae}, optical modulators\cite{liu}, plasmonic devices\cite{tim}, microcavities\cite{engel}, and ultrafast lasers\cite{sun}, just to cite a few. Understanding the interaction of light with graphene, which in the first instance creates optically excited (``hot") carriers, is pivotal to all these optoelectronic applications.

The interaction of light with carriers creates an out-of-equilibrium distribution, which relaxes on an ultrafast timescale to a hot Fermi-Dirac distribution, that subsequently cools via phonon emission. While the slower relaxation mechanisms have been extensively investigated\cite{Norris2004,Obr2011,Huang2010}, the initial stages of relaxation, ruled by fundamental electron-electron (e-e) interactions\cite{gonzalez_prl_1996, gonzalez_prl_1999, hwang_prb_2007,polini_prb_2008}, still pose a challenge. Experimentally, they defy the resolution of most pump-probe setups, due to the ultrafast sub-100-fs dynamics spanning a broad range of energies. Theoretically, the linear dispersion of massless Dirac fermions poses a novel many-body problem, fundamentally different from the parabolic-band model used for decades in ordinary metals and semiconductors\cite{Pines_and_Nozieres,Giuliani_and_Vignale}.

\begin{figure*}
\centerline{\includegraphics[width=0.7\linewidth]{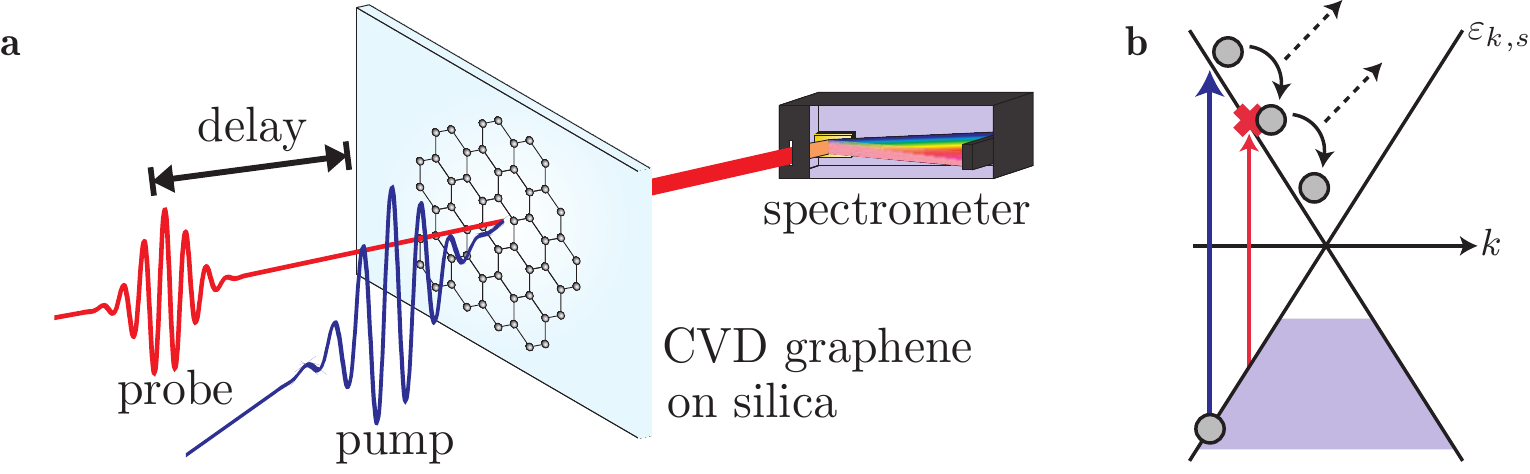}}
\caption{a) Schematic of pump-probe experiment: a pump and a probe pulse, with different colors, impinge on the sample with a variable delay. The transmission of the probe pulse through the sample is measured by a detector. b) Massless Dirac fermion bands in graphene, $\varepsilon_{{\bm k}, s} = s\hbar v |{\bm k}|$ with $s = \pm 1$. Pump (blue arrow) and probe pulses (red arrow) are applied at different photon energies in a hole doped sample. The electron distribution relaxes towards the Dirac point by losing energy. Electrons (grey circles) transfer energy to other degrees of freedom (dashed arrows), such as phonons and other electrons in the occupied Fermi sea. When half of the probe-pulse energy matches the maximum of the electron distribution, absorption is strongly suppressed by Pauli blocking (red cross) and the transition is ``bleached".\label{fig:one}}
\end{figure*}

The non-equilibrium dynamics of hot carriers can be very effectively studied by ultrafast pump-probe spectroscopy. In this technique an ultrashort laser pulse creates a strongly out-of-equilibrium (non-thermal) distribution of electrons in conduction band and holes in valence band. Optically-excited carriers relax, eventually reaching thermal equilibrium with the lattice. The relaxation dynamics, due to various scattering processes, including e-e and electron-phonon (e-ph) scattering, as well as radiative electron-hole (e-h) recombination, is then accessed by a second probe pulse (see Fig.\ref{fig:one}). The time-evolving distribution of hot electrons inhibits the absorption of probe light due to Pauli blocking, yielding an increase in the transmission through the sample (``photobleaching", PB) which is best probed at longer wavelength (lower energy) with respect to the excitation pulse. Transient absorption thus enables the direct measurement of the distribution function in real time. A typical time evolution of the hot-electron distribution, calculated via the microscopic theory discussed later, is shown in Fig.\ref{fig:five}e. Here we are interested in the sub-$1~{\rm ps}$ dynamics during which two main processes occur. Firstly, the initial peak produced by the pump laser broadens due to e-e collisions converging towards a hot Fermi-Dirac shape in an ultrashort time scale\cite{gonzalez_prl_1996, gonzalez_prl_1999, hwang_prb_2007,polini_prb_2008} (less than $50~{\rm fs}$). Subsequently, the optical phonons emission\cite{lazzeri} drives a cooling process in which the peak of the Fermi-Dirac distribution shifts to lower energies towards the Dirac point.

In order to experimentally access the very first stages of relaxation and to disentangle the role of e-e scattering from other scattering mechanisms, it is therefore necessary to probe low-energy electronic transitions {\it and} at the same time achieve a sub-$10~{\rm fs}$ temporal resolution. Pump-probe spectroscopy has been extensively employed to study relaxation processes in carbon-based materials. A variety of different samples have been studied, including thin graphite/multilayer graphene flakes\cite{Breus2009,Obr2011,Carbone2011}, few-layer graphene sheets on SiC\cite{Norris2004,Dawlaty2008,Huang2010} and graphene oxide\cite{Ruzicka2010,JPC2009}, but only a limited number of studies reported experiments on Single-Layer Graphene (SLG)\cite{Hale2011,Heinz2010,BKWMM2011}. While the slower relaxation mechanisms have been extensively investigated, the initial stages of relaxation, ruled by fundamental e-e interactions still pose a challenge. Experimentally, they defy the resolution of most pump-probe setups, due to the extremely fast sub-100-fs carrier dynamics spanning a broad range of energies. Theoretically, the linear dispersion of massless Dirac fermions poses a novel many-body problem, fundamentally different from the parabolic-band model used for decades in ordinary metals and semiconductors. Indeed, the temporal resolution reported in earlier literature, either in degenerate or two-color pump-probe, was limited to$\sim$ 100fs or higher\cite{Norris2004,Obr2011,Huang2010,Hale2011,Dawlaty2008,Ruzicka2010}. This prevented the direct observation of the intrinsically fast e-e scattering processes. Earlier studies thus mostly targeted the phonon-mediated cooling of a thermalised (but still hot) electron distribution, established within the pump pulse duration. To date, only Ref.\onlinecite{BKWMM2011} reported the investigation of SLG with sub-$10~{\rm fs}$ time resolution. However, this particular experiment was conducted in a degenerate scheme, thus providing only a limited access to the electron relaxation dynamics.

The focus of this work is on the impact of e-e interactions on the initial stages of the non-equilibrium dynamics. Even at equilibrium, e-e interactions are responsible for a wealth of exotic phenomena in graphene\cite{kotov_rmp_2012}. They reshape the Dirac bands\cite{elias_naturephys_2011,bostwick_science_2010} and substantially enhance the quasi-particle velocity\cite{elias_naturephys_2011}. Angle-resolved photoemission spectroscopy showed electron-plasmon interactions in doped samples\cite{bostwick_naturephys_2007,bostwick_science_2010}, and a marginal Fermi-liquid behavior in undoped ones\cite{siegel_pnas_2011}. Many-body effects were also revealed in optical spectra both in the infrared (IR)\cite{li_naturephys_2008} and in the ultraviolet, where strong excitonic effects were measured\cite{mak_prl_2011,kravets_prb_2010}. In the non-equilibrium regime, the extremely fast e-e relaxation occurring on the time scale of tens of femtoseconds is also consistent with the theoretical results of Refs.\onlinecite{butscher_apl_2007,MWBK2011, kim_prb_2011,sun_prb_2012}. However, these pioneering approaches\cite{butscher_apl_2007,MWBK2011,kim_prb_2011,sun_prb_2012} relied solely on numerical methods and, as discussed below, did not take full advantage of the symmetries of the scattering problem. This implies an uncontrolled treatment of crucially important scattering events, which are collinear, thus characterized by a high degree of symmetry. A deeper theoretical understanding of collinear scattering events and, most importantly, their phase space, requires more analytical work.

Here we combine extreme temporal resolution broadband pump-probe spectroscopy with a microscopic semi-analytical theory based on the quantum Boltzmann equation to investigate e-e collisions in graphene during the very early stages of relaxation. We identify the fundamental processes controlling the ultrafast dynamics in graphene, in particular the significant role of Auger processes, including charge multiplication.
\begin{figure*}
\centerline{\includegraphics[width=0.7\linewidth]{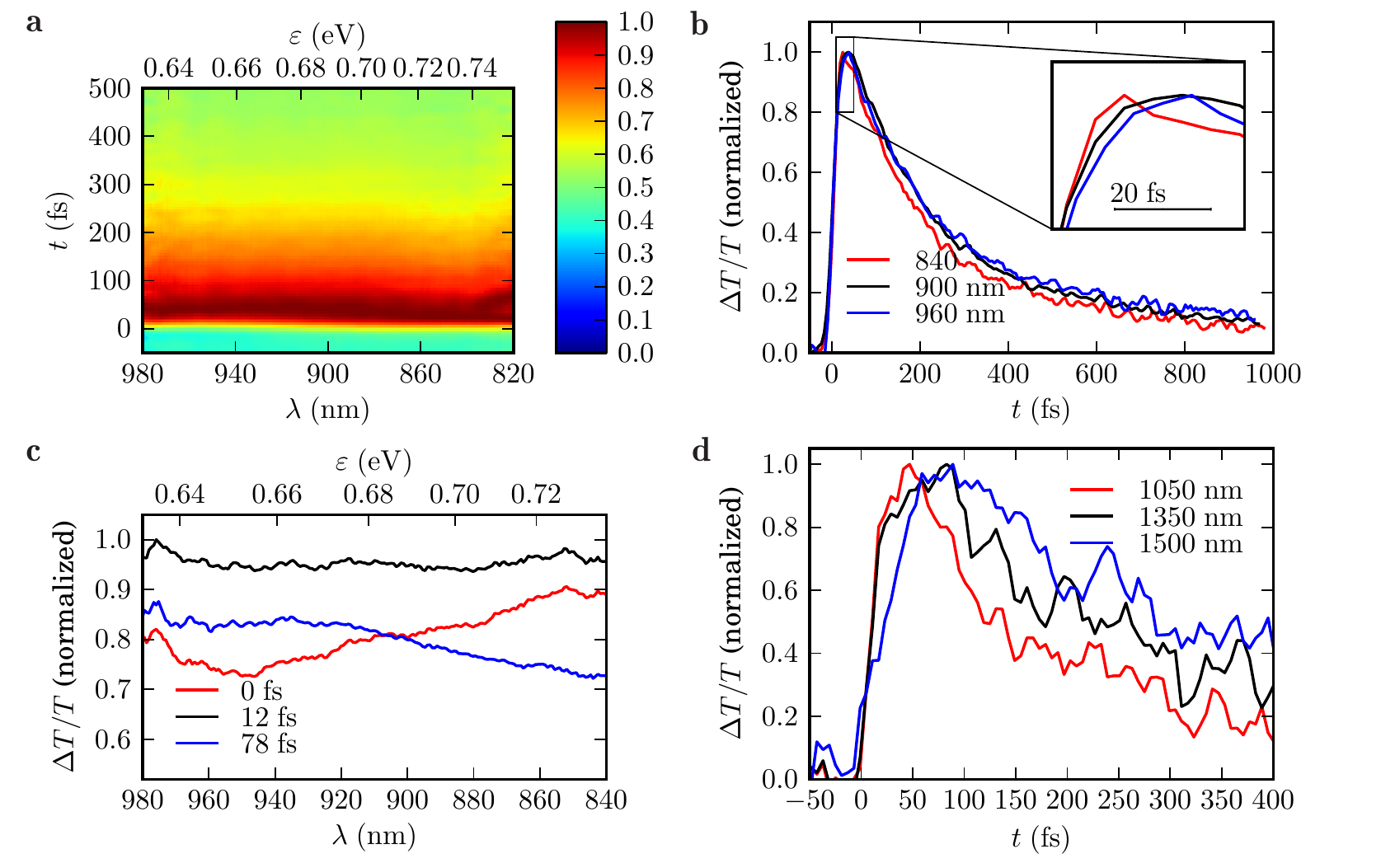}}
\caption{a) $\Delta T/T$ map as a function of probe wavelength $\lambda$ and pump-probe delay $t$. The positive photo-bleaching signal due to Pauli blocking rises on the $10~{\rm fs}$ timescale due to ultrafast spreading of the e-distribution upon impulsive excitation. b) Transient dynamics at selected pump-probe map $\lambda$s. The inset shows that the onset of the signal moves to longer times as the probe photon energy decreases. c) Transient spectra at selected delays. For delays below $20~{\rm fs}$ the signal peaks at high photon energies due to the strongly out-of-equilibrium electron distribution. At later ($\sim 20~{\rm fs}$) times the signal flattens and then peaks at low photon energies, as the electron distribution thermalizes. d) Transient dynamics at probe photon energies $< 1~{\rm eV}$. The delay in the photobleaching onset is more evident. The recovery time slows as the e-distribution approaches the Dirac point
\label{fig:two}}
\end{figure*}

SLG is grown by Chemical Vapor Deposition (CVD)\cite{Li_s_2009,bae} and transferred onto 100${\rm \mu m}$ quartz substrates, as described in Methods. These substrates induce negligible artifacts in the experiments, as verified by measuring the blank. We perform two-color pump-probe spectroscopy using few-optical-cycle pulses (see Methods). We impulsively excite inter-band transitions with a $7~{\rm fs}$ pulse at $2.25~{\rm eV}$ ($2$-$2.5~{\rm eV}$ bandwidth) and probe with a red-shifted $13~{\rm fs}$ pulse ($1.2$-$1.45~{\rm eV}$ bandwidth), as well as a $9~{\rm fs}$ pulse ($0.7$-$1.2~{\rm eV}$ bandwidth). The density of photoexcited electrons is$\sim 10^{13}~{\rm cm}^{-2}$. The availability of such short IR pulses allows us to follow the electron population as it evolves towards a Fermi-Dirac distribution. Our instrumental response function (IRF) (full width at half maximum of the pump-probe cross-correlation) is less than $15~{\rm fs}$\cite{MPC2006,BMCMB2010}, with a crucially important order of magnitude improvement in time resolution with respect to previous two-color studies\cite{Ruzicka2010,Huang2010,Obr2011}. This allows us to directly probe the e-e dynamics, unlike previous works.

Fig.\ref{fig:two}a plots the two dimensional (2D) map of the differential transmission ($\Delta T / T$) spectra as a function of pump-probe delay in the $1.2$-$1.45~{\rm eV}$ spectral range. We observe, even with our time resolution, an almost pulsewidth-limited rise of the PB signal in the near-IR, Fig.\ref{fig:two}b. This immediately points to an ultrafast e-e relaxation, taking place over a timescale comparable to our IRF. The PB signature is nearly featureless as a function of probe wavelength, as expected given the linear dispersion of massless Dirac fermions (MDFs) in SLG. The selected time traces at different probe photon energies feature a biexponential decay, with a first time constant $\tau_{1}\simeq 150-170~{\rm fs}$, and a second longer time constant $\tau_{2} >1~{\rm ps}$. In agreement with previous studies, we assign the first decay to the cooling of the hot electron distribution via interaction with optical phonons, and the longer decay to relaxation of the thermalized electron and phonon distributions by anharmonic decay of hot phonons\cite{Kaindl2009,VanDriel2009}. By varying the excitation intensity we observe a linear dependence of the PB peak on pump fluence, while its dynamics is nearly fluence-independent.

A deeper insight into the e-e thermalization process can be obtained from the inset of Fig.\ref{fig:two}b, showing a delay in the PB maximum onset at longer probe wavelengths. In addition, by comparing $\Delta T / T$ at selected probe delays (Fig.\ref{fig:two}c) we see that, at early times ($\sim 10~{\rm fs}$), $\Delta T / T$ has a positive slope, peaking at high photon energy. Starting from $\sim 20~{\rm fs}$, it progressively flattens and changes to a negative slope, which persists and increases at longer delays. To understand the data, we recall that $\Delta T/T$ is proportional to the transient electron distribution\cite{Obr2011,BKWMM2011} at time $t$. The sub-$10~{\rm fs}$ $2.25~{\rm eV}$ pump pulse creates an electron distribution peaking at$\simeq 1.12~{\rm eV}$ above the Fermi level (red line in Fig.\ref{fig:five}e), while the probe pulse samples the $0.6$-$0.72~{\rm eV}$ interval. At early times we therefore observe the tail of this distribution, with a positive slope. On the other hand, a thermal Fermi-Dirac distribution, even with a typical doping usually present in as-prepared SLG (chemical potential$\sim 100$-$200~{\rm meV}$)\cite{Casiraghi_apl_2007}, peaks at low photon energies, yielding a differential transmission with a negative slope. The transition from the non-thermal to the thermal regime, which is completed within$\sim 50~{\rm fs}$, is responsible for the change of slope in $\Delta T/T$. Fig.\ref{fig:two}d plots the $\Delta T / T$ dynamics measured with the second red-shifted IR probe pulse. The high temporal resolution combined with the low photon energy allows us to observe an even clearer delay in the PB peak formation. In particular, the maximum $\Delta T / T$ is reached after $40$, $60$, and~$80~{\rm fs}$ for probe wavelengths of $1050~{\rm nm}$ ($1.2~{\rm eV}$), $1350~{\rm nm}$ ($0.92~{\rm eV}$), and~$1550~{\rm nm}$ ($0.8~{\rm eV}$), respectively. To the best of our knowledge, this is the first experiment setting a timescale for the out-of equilibrium carrier thermalization with a direct measurement. Tuning the probe to longer wavelengths also allows us to follow the subsequent e-ph cooling of the carrier distribution. In fact, $\tau_{1}$ becomes longer ($\simeq 400~{\rm fs}$ at $1550~{\rm nm}$) when probing at smaller photon energies, consistent with a distribution moving towards the Dirac point, before dissipating the excess energy into the phonon bath (Fig.\ref{fig:five}e).
\begin{figure*}
\centerline{\includegraphics[width=0.7\linewidth]{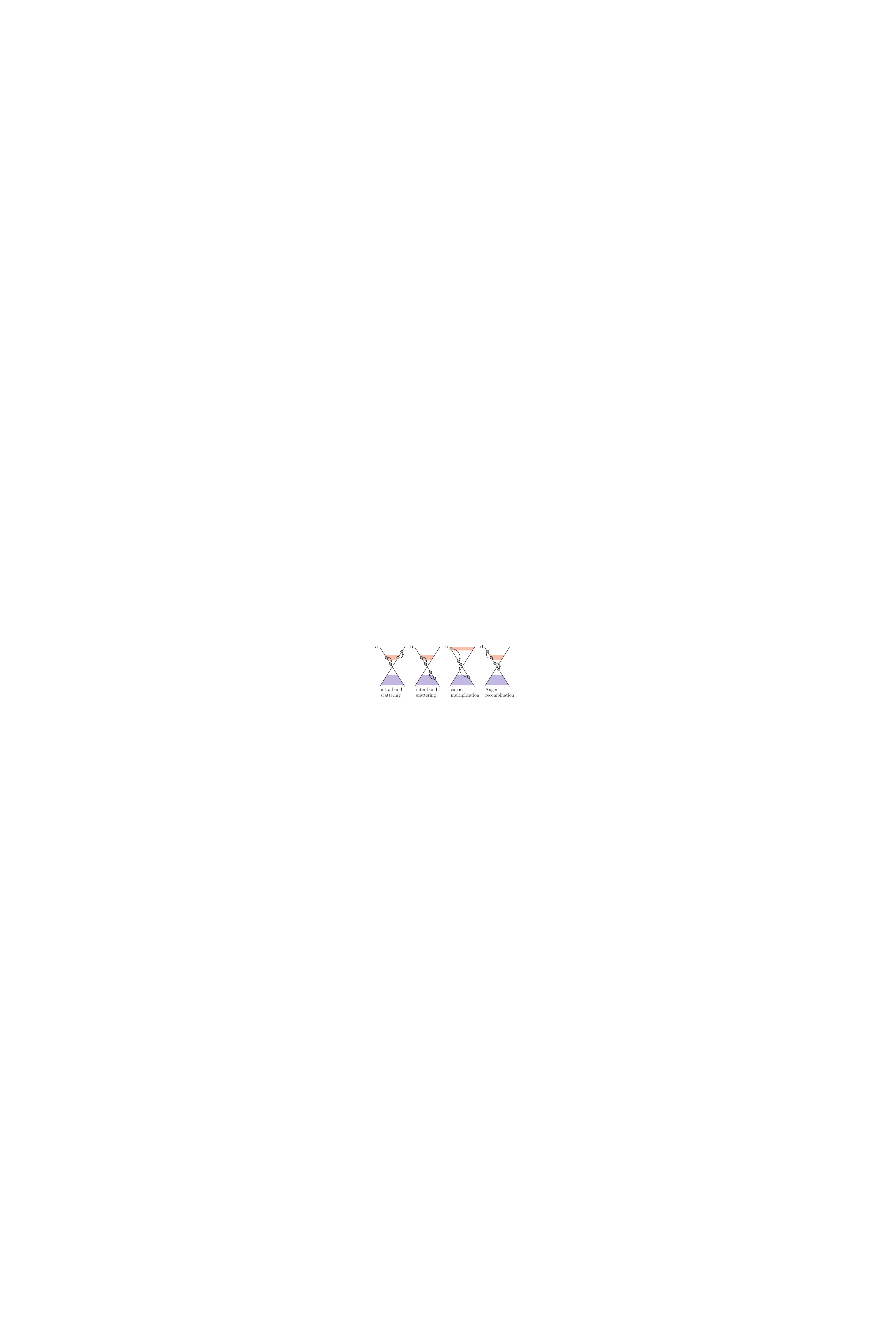}}
\caption{Coulomb-enabled two-body scattering processes in graphene. Shaded areas denote occupied states, in conduction and valence band, in a non-equilibrium state, at a given time. Arrows mark transitions from initial to final states. Coulomb collisions can take place between electrons in the same band [(a) intra-band scattering] or between electrons in different bands [(b) inter-band scattering]. Electrons can also scatter from one band to the other (c,d). These ``Auger processes" can either {\it increase} [(c) carrier multiplication] or {\it decrease} [(d) Auger recombination] the conduction-band electrons\label{fig:three}}
\end{figure*}

\begin{figure*}
\centerline{\includegraphics[width=0.7\linewidth]{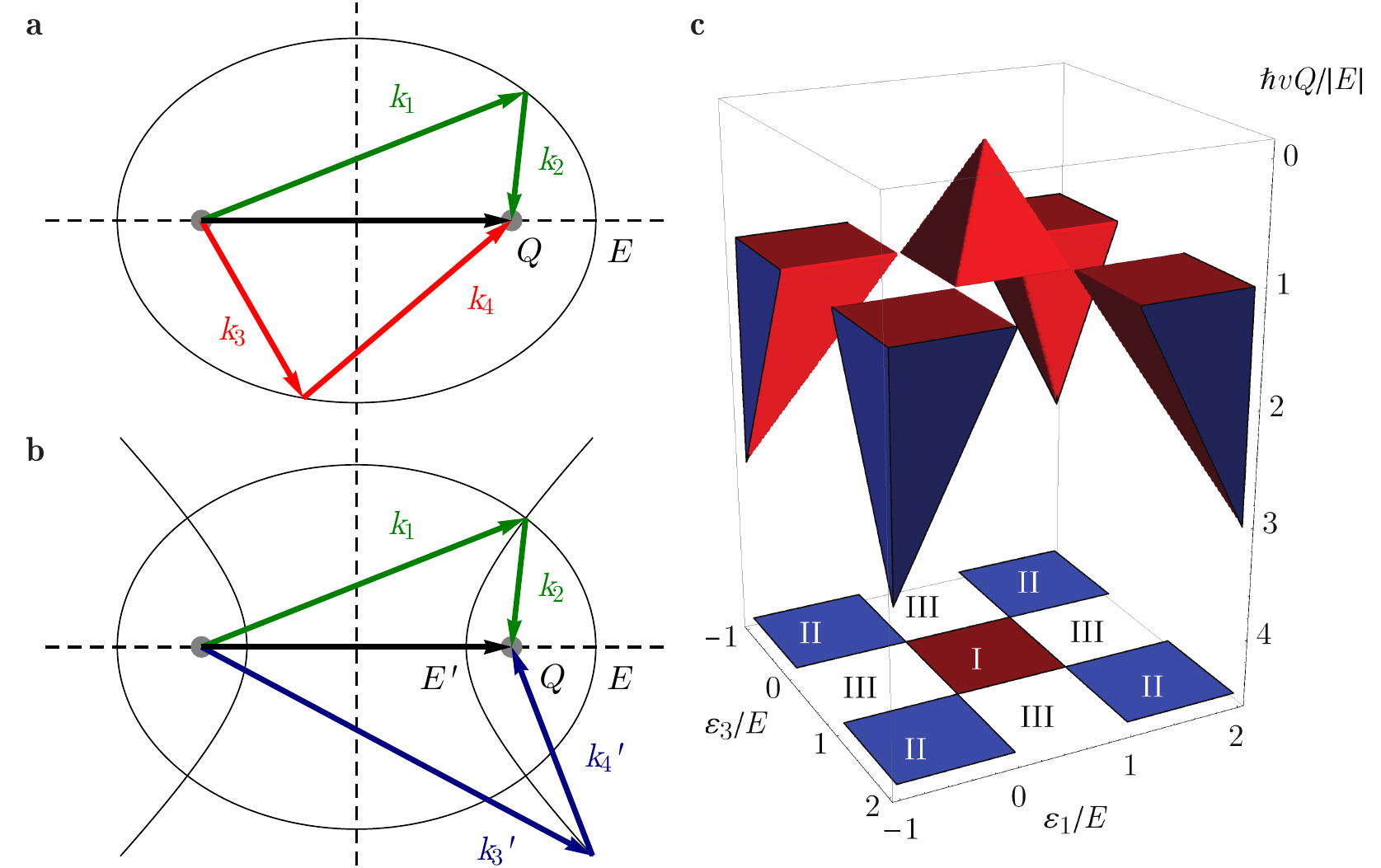}}
\caption{Phase space for two-body collisions in graphene. a) Allowed and b) forbidden scattering process in which two electrons with momenta ${\bm k}_{1}$ and ${\bm k}_{2}$ scatter into momenta ${\bm k}_3$ and ${\bm k}_4$ (${\bm k}_{3}'$ and ${\bm k}_{4}'$). The total momentum ${\bm Q}={\bm k}_1+{\bm k}_2$ is conserved in both panels. Intra-band (inter-band) scattering can be represented on an ellipse (hyperbola) with distance between foci equal to $\hbar v |{\bm Q}|$, where $v$ is the Fermi velocity, and the major axis equals the total energy $E$ ($E'$). The total energy $E=\hbar v(|{\bm k}_1|+|{\bm k}_2|)$ is conserved in a), which represents intra-band scattering. On the contrary, in b), $E'=\hbar v(|{\bm k}'_3|-|{\bm k}'_4|)$ of the outgoing particles is {\it smaller} than $E$. This forbidden scattering process represents a collision between two incoming particles in the same band, and two outgoing particles in different bands, {\it i.e.} an Auger process, for Figs.\ref{fig:three}c,d. Energy conservation $E = E'$ implies that Auger processes can only take place in the ``degenerate limit", {\it i.e.} when the vertices of the two confocal conical sections coincide with the foci. In this limit the ellipse and hyperbola collapse onto a segment and half-line, respectively, and all the momenta are {\it collinear}. c) The three-dimensional solid represents the allowed values of the total momentum $|{\bm Q}|$ [in units of $E/(\hbar v)$] plotted as a function of energies $\varepsilon_1$ and $\varepsilon_3$ (both in units of the total energy $E$) of an incoming and outgoing particle, respectively. Region I corresponds to intra-band processes; region II to inter-band processes; region III to Auger processes. No phase space is available for carrier multiplication and Auger recombination for massless Dirac fermions in 2D
\label{fig:four}}
\end{figure*}

The ultrashort time needed for a hot electron distribution to thermalize in SLG is a consequence of e-e collisions~\cite{gonzalez_prl_1996, gonzalez_prl_1999, hwang_prb_2007,polini_prb_2008,butscher_apl_2007,MWBK2011,sun_prb_2012}. We now proceed to discuss theoretically Coulomb-mediated two-body collisions in SLG (Fig.\ref{fig:three}). These include intra- and inter-band scattering, ``impact ionization" or ``carrier multiplication" (CM), and Auger recombination. In a CM process, for example, electrons in valence band are ``ejected" from the Fermi sea and promoted to unoccupied states in conduction band. CM in graphene could thus play a pivotal role in the realization of very efficient photovoltaic devices and photodetectors with ultra-high sensitivity\cite{WKM2010,rana_prb_2007,girdhar_apl_2011,winzer_prb_2012}. Ref.\onlinecite{rana_prb_2007} noted that, due to severe kinematic constraints (see Figs.\ref{fig:four}a,b) for MDFs in 2D, these processes can only take place in a {\it collinear} scattering configuration. To the best of our knowledge, however, it has not yet been shown that CM and Auger recombination can occur in a 1d manifold embedded in 2D space, since the incoming and outgoing momenta of the scattering particles lie on the same line. Intuitively these processes should thus be nearly irrelevant. Indeed, in Fig.\ref{fig:four}c we demonstrate that the phase space for CM and Auger recombination in 2D MDF bands {\it vanishes}. We will come back to this issue below, in connection with the theoretical calculation of the e-e contribution to the collision integral in the quantum Boltzmann equation (QBE).

We use this knowledge of Coulomb-mediated collisions in the framework of the QBE, the standard tool to investigate electron dynamics in metals and semiconductors\cite{HaugJauhoBook,snoke_annphys_2011}. We consider the equations of motion for the electron ($f_{s,\mu}({\bm k})$) and phonon ($n_{\bm q}^{(\nu)}$) distributions, including transverse and longitudinal optical phonon modes at the ${\bf \Gamma}$ and ${\bf K}$ points of the Brillouin zone\cite{PLMFR2004,lazzeri_prb_2008,basko_prb_2009}. Here $s = +1$ ($s = -1$) labels the conduction (valence) band and the valley degree-of-freedom, $\mu = \pm 1$ indicates whether the electron wavevector ${\bm k}$ is measured from ${\bf K}$ or ${\bf K}'$. The electron distribution is independent of the spin degree-of-freedom. QBE describes i) Coulomb scattering between electrons and ii) phonon-induced electronic transitions, where the energy of an electron decreases (increases) by the emission (absorption) of a phonon. In particular, the phonons at ${\bf \Gamma}$ are responsible for intra-valley transitions, while inter-valley transitions involve ${\bf K}$ phonons. Emission of optical phonons is crucial in the cooling stage of the dynamics and is possible because the energy of photoexcited electrons ($\sim 1~{\rm eV}$) is much higher than the typical phonon energy ($\sim 150$-$200~{\rm meV}$)~\cite{PLMFR2004}. Finally, ph-ph interactions, arising from the anharmonicity of the lattice, are taken into account phenomenologically\cite{MWBK2011,sun_prb_2012}, by employing a linear relaxation term, parametrised by a decay rate $\gamma_{\rm ph}/\hbar$.

We neglect acoustic phonons, since they are expected to modify the electron dynamics on a $>1~{\rm ps}$ timescale\cite{bistritzer_prl_2009,tse_prb_2009,Kaindl2009,VanDriel2009}. Even when scattering between electrons and acoustic phonons is assisted by disorder, the so-called ``supercollision" process\cite{song_arXiv_2011}, relaxation times$\sim 1$-$10~{\rm ps}$ have been predicted\cite{song_arXiv_2011} and observed\cite{graham_arXiv_2012}. These are still too long compared to those considered here.

To simulate the experiments, we solve the QBE with an initial condition given by the superposition of a Fermi-Dirac distribution in equilibrium with the lattice at $T = 300~{\rm K}$ and a Gaussian peak (dip) in conduction (valence) band, centered at $\bar{\varepsilon} = \pm 1.125 ~{\rm eV}$, with a width of $0.09~{\rm eV}$. Ref.\onlinecite{MWBK2011} argued that the anisotropy introduced in the electron distribution by the pump pulse disappears in a few fs due to e-e scattering. We thus enforce circular symmetry in our QBE to deal with time-dependent distribution functions, $f_\mu(\varepsilon_{{\bm k}, s})$, not dependent on the polar angle $\theta_{\bm k}$ of ${\bm k}$, but on $|{\bm k}|$ and $s = \pm 1$ only, through the Dirac-band energy $\varepsilon_{{\bm k}, s}= s \hbar v |{\bm k}|$, where $v \approx 10^6~{\rm m}/{\rm s}$ is the Fermi velocity. Thus, the e-e contribution to the QBE for the electron distribution can be written as:
\begin{widetext}
\begin{eqnarray}\label{eq:circularQBE}
\left . \frac{d f_{\mu}(\varepsilon_{1})}{dt} \right |_{\rm e-e} = \int_{-\infty}^{\infty} d\varepsilon_{2} \int_{-\infty}^{\infty} d\varepsilon_{3}~{\cal C}_{\mu}(\varepsilon_{1},\varepsilon_{2},\varepsilon_{3})\lbrace [1 - f_{\mu}(\varepsilon_{1})] [1 - f_{\mu}(\varepsilon_{2})]f_{\mu}(\varepsilon_{3}) f_{\mu}(\varepsilon_4) - & & \nonumber \\
f_{\mu}(\varepsilon_{1}) f_{\mu}(\varepsilon_{2}) [1 -  f_{\mu}(\varepsilon_{3})][1 - f_{\mu}(\varepsilon_{4})] \rbrace~, & &
\end{eqnarray}
\end{widetext}
where ${\cal C}_\mu(\varepsilon_{1},\varepsilon_{2},\varepsilon_{3})$ is the Coulomb kernel (see Methods) describing the exchange of (momentum and) energy from
$\varepsilon_{1}$ and $\varepsilon_{2}$ (incoming states) to $\varepsilon_{3}$ and $\varepsilon_{4} = \varepsilon_{1} + \varepsilon_{2} - \varepsilon_{3}$ (outgoing states) during a two-body (intra-valley) collision. Conservation of energy and momentum are automatically enforced in Eq.(\ref{eq:circularQBE}).

Circular symmetry allows us to treat the angular integrations in the Coulomb kernel analytically, taking particular care of the contributions arising from the subtle collinear scattering processes described above (see Methods). The contribution of intra- and inter-band processes can then be cast into an integration over the allowed total momentum ${\bm Q}$ (see Fig.\ref{fig:four}c).

CM and Auger recombination require additional care. As described in Fig.\ref{fig:four}c, their phase space in the case of 2D MDFs {\it vanishes} if momentum and energy are conserved. This statement holds true for infinitely-sharp bare bands with strictly linear dispersions. Non-linear corrections to the MDF Hamiltonian in powers of momentum (measured from the Dirac point), however, appear due to lattice effects ({\it e.g.} trigonal warping). These are small in the range of energies set by the pump ($\sim 1~{\rm eV}$)\cite{castroneto_rmp_2009}. Non-linearities appear also due to the inclusion of e-e interactions\cite{kotov_rmp_2012}. These give rise to a self-energy correction to the bare MDF bands, whose real part is responsible for the velocity enhancement\cite{elias_naturephys_2011}. This correction becomes significantly large in the low-doping regime\cite{kotov_rmp_2012,elias_naturephys_2011}, while here we are interested in the non-equilibrium dynamics of a substantial population of photexcited electrons ($\sim 10^{13}~{\rm cm}^{-2}$). Most importantly, any effect giving a finite width to the quasiparticle spectral function, such as e-e interactions\cite{bostwick_naturephys_2007,bostwick_science_2010}, opens up a phase space for Auger processes. To calculate the Auger contribution to ${\cal C}_\mu(\varepsilon_{1},\varepsilon_{2},\varepsilon_{3})$ we take into account these electron-lifetime effects by a suitable limiting procedure (see Methods). We stress that the final result is {\it independent} of the precise mechanism limiting the electron lifetime.

Crucially, we go beyond the Fermi golden rule\cite{Pines_and_Nozieres,Giuliani_and_Vignale} by including screening in the matrix element of the Coulomb interaction, by using the Random Phase Approximation (RPA)\cite{Pines_and_Nozieres,Giuliani_and_Vignale}. To this end, we introduce the screened potential\cite{Giuliani_and_Vignale} $W(q, \omega; t) = v_q/\epsilon(q,\omega; t)$, where $q$ and $\hbar\omega$ are the momentum and energy transfer in a scattering event, respectively, $v_q = 2\pi e^2/({\bar \epsilon} q)$ is the 2D Fourier transform of the Coulomb potential, ${\bar \epsilon}$ is an average dielectric screening, depending on the media around the sample\cite{kotov_rmp_2012}. The RPA dynamical dielectric function is $\epsilon(q,\omega; t)=1- v_q \chi^{(0)}(q,\omega; t)$, where the non-interacting {\it time-dependent} polarization function $\chi^{(0)}(q,\omega; t)$ depends on the distribution function $f_\mu$ at time $t$:
\ber\label{lehmanrepresentation}
\chi^{(0)}(q,\omega; t) &=& 2 \sum_{s, s', \mu}\int \frac{d^2{\bm k}}{(2\pi)^2}\frac{f_\mu(\varepsilon_{{\bm k}, s}) - f_\mu(\varepsilon_{{\bm k} +{\bm q}, s'})}{\hbar \omega + \varepsilon_{{\bm k}, s} - \varepsilon_{{\bm k} +{\bm q}, s'}+i0^+}\nonumber\\
&\times&|F^{(\mu)}_{s s'}(\theta_{\bm k} - \theta_{{\bm k} + {\bm q}})|^2~.
\eer
Here the factor two accounts for spin degeneracy, and the chirality factor $F^{(\mu)}_{ss'}$, which depends on the polar angle $\theta_{\bm k}$ of ${\bm k}$, is defined in Methods.

Collinear scattering plays also a key role in the theory of screening of 2D MDFs. It takes place on the ``light cone" $\omega=vq$ when ${\bm k}+{\bm q}$ is either parallel or anti-parallel to ${\bm k}$ in Eq.(\ref{lehmanrepresentation}). This implies a strong peak in the imaginary part of $\chi^{(0)}(q,\omega; t)$ (which physically represents the spectral density of particle-hole pairs) close to the light cone\cite{kotov_rmp_2012}, where $\Im m~[\chi^{(0)}(q,\omega; t)]$ diverges like $|\omega^2 - v^2q^2|^{-1/2}$: RPA dynamical screening {\it suppresses} Auger scattering. Since we are looking at effects that are very fast on the time-scale set by plasma oscillations ($1$-$10~{\rm THz}$)\cite{polini_prb_2008}, we also introduce a ``static" approximation. This consists in evaluating $\chi^{(0)}(q,\omega; t)$ at $\omega$=0. In the static limit there is no collinear contribution to the screened potential and the impact of Auger processes is maximal. Note that $\chi^{(0)}(q,0; t)$ still depends on time through $f_\mu$. RPA is definitely a very good starting point to deal with screening in metals and semiconductors\cite{Pines_and_Nozieres,Giuliani_and_Vignale}, but is certainly not exact. We are thus allowed to modify the RPA dynamical screening function to interpolate the strength of Auger processes between its maximal (static screening) and minimal (dynamic screening) value. We thus introduce a third approximate screening model by cutting off the singularity of $\chi^{(0)}(q,\omega; t)$ in the region $|\hbar(\omega - vq)| \leq \Lambda$ of width $2\Lambda$ near the light cone. This regularized polarization function, $\chi^{(0)}_\Lambda(q,\omega; t)$, leads to a regularized screened potential $W_\Lambda(q,\omega;t)= v_q/\varepsilon_\Lambda(q,\omega;t) \equiv v_q/[1-v_q \chi^{(0)}_\Lambda(q,\omega; t)]$, with $\Lambda = 20~{\rm meV}$ in our calculations.

We stress that our theory is free of fitting parameters and it is predictive from the IR to the optical domain. In the regularized screening model we do {\it not} adjust the value of $\Lambda$ to yield the best agreement with experiments. However, the theory is expected to work better in the IR limit, since it is based on the low-energy MDF Hamiltonian and thus neglects band-structure effects, which become non-negligible at high energy.
\begin{figure*}
\centerline{\includegraphics[width=0.7\linewidth]{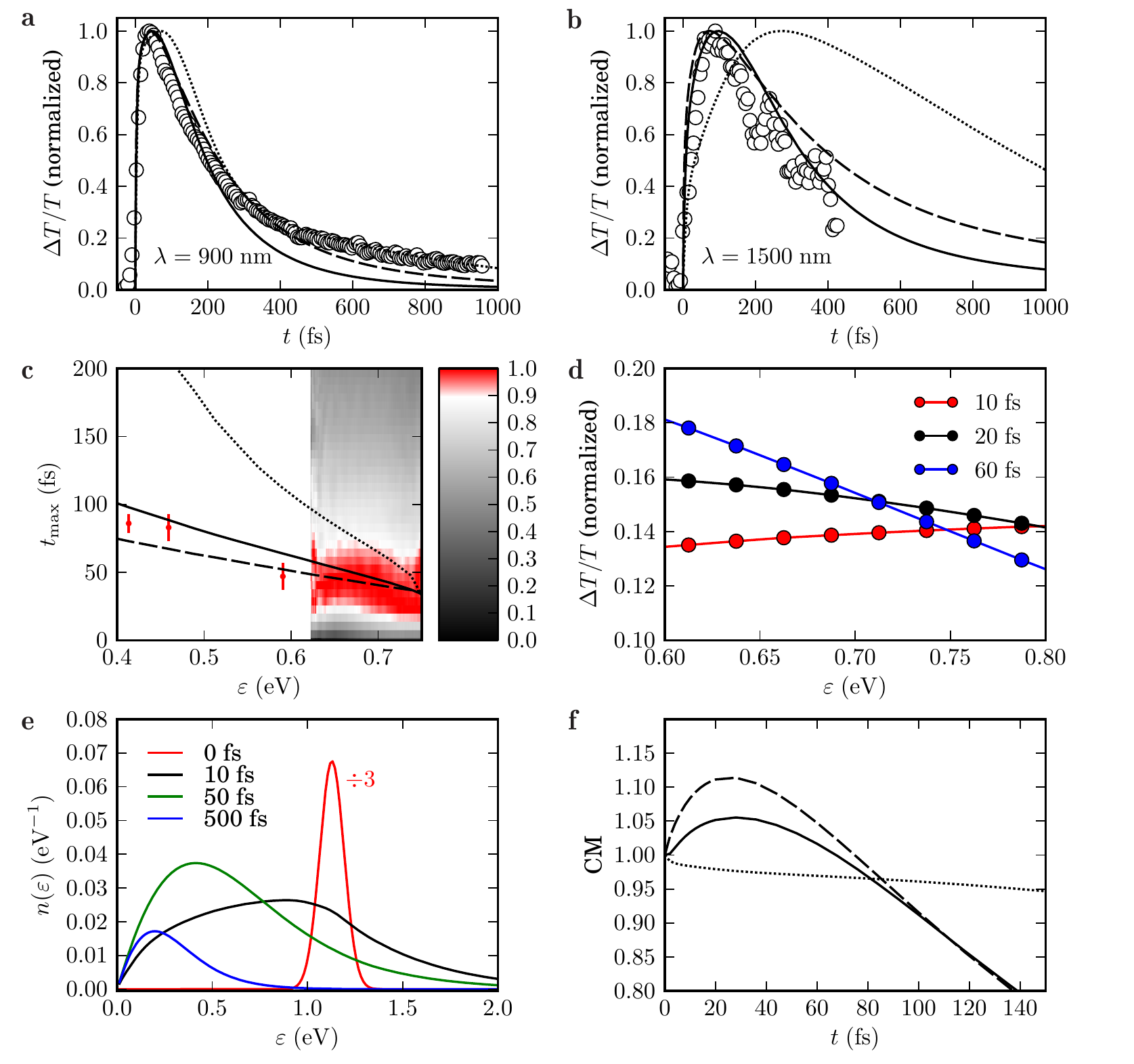}}
\caption{a) Time-evolution of $\Delta T/T$ at $\lambda \simeq 900~{\rm nm}$ extracted from Fig.\ref{fig:two} (circles). Theoretical results are obtained from the QBE solution using dynamical screening (dotted line), regularized dynamical screening (solid line), and static screening (dashed line). All data are normalized to their maximum, and correspond to a chemical potential$\sim$200meV (hole-doped sample). b) As in panel a) for $\lambda \simeq 1500~{\rm nm}$. c) Time $t_{\rm max}$ (in ${\rm fs}$) at which $\Delta T/T$ reaches its maximum. The labeling of the theoretical data (lines) is the same as in a,b). Experimental data in Fig.\ref{fig:two}a are here shown as a color plot in a continuous optical spectral range for $\lambda \lesssim 1000~{\rm nm}$. The circles with error bars correspond to three IR measurements. d) $\Delta T/T$ as a function of the electron energy $\varepsilon$ at different times. The slope inversion, signature of the initial stage of the dynamics (cfr. Fig.\ref{fig:two}c), is correctly reproduced by the theory. e) Time-evolution of the electron population $n(\varepsilon)$ per unit cell (in units of ${\rm eV}^{-1}$) as derived by solving the QBE with regularized dynamical screening. The initial hot- electron peak (red) is centered at half the energy of the pump laser. The amplitude of this distribution is divided by $3$ to fit into the frame of the figure. The hot-electron peak rapidly broadens into a non-thermal distribution (black), which then thermalizes to a hot Fermi-Dirac (green) distribution. Subsequently, cooling by phonon emission takes place (blue), until thermal equilibrium with the lattice is eventually established (not shown here since other effects neglected in our theory, such as acoustic phonons, are important in the late stages of the dynamics. When the $n(\varepsilon)$ peak energy crosses half the energy of the probe, Pauli blocking inhibits absorption. In this case a stronger transmitted ``bleaching'' signal is recorded at the detector. f) Carrier multiplication as a function of time. Labeling as in panels a-c). Note the suppression of carrier multiplication in the prediction based on dynamical screening (dotted line)\label{fig:five}}
\end{figure*}

Fig.~\ref{fig:five} shows that the theory with dynamical screening compares poorly with experiments, in predicting both the prompt PB onset and its subsequent decay. While with static and regularized dynamical screening the profile of $\Delta T/T$ is in good agreement with experiments (Figs.\ref{fig:five}a,b), the dynamics is much slower in the presence of dynamical screening. This is best seen at low probe energies, Fig.\ref{fig:five}b. The dependence of the maximum of the differential transmission (Fig.\ref{fig:five}c) on probe wavelength further highlights the large discrepancy between experiments and theory with dynamical screening. We trace back this discrepancy to the fact that, as stated above, dynamical screening completely {\it suppresses} Auger processes. We thus conclude that these processes are a crucially important relaxation channel for the non-equilibrium electron dynamics in graphene. Fig.\ref{fig:five}c shows clearly that it would have not been possible to draw this conclusion without comparing theoretical predictions with experimental data in the low-energy regime, {\it i.e.} for energies $\varepsilon < 0.6~{\rm eV}$.

A closer inspection of the dynamics reveals that the thermalization of the initial hot electron distribution (see Fig.~\ref{fig:five}e) is accompanied by a very fast equilibration of the chemical potentials of conduction and valence bands, over few tens of fs. Auger processes are the only e-e scattering channel that couples the two bands (see Figs.\ref{fig:three}c,d) and are thus responsible for this ultrafast equilibration (e-ph scattering is relevant on the much longer time scale of hundreds of fs\cite{Norris2004,Dawlaty2008}). Thus, hot carriers in conduction band provide a substantial amount of energy for the promotion of electrons from valence to conduction band, resulting in CM\cite{WKM2010,rana_prb_2007,girdhar_apl_2011,winzer_prb_2012}. This is shown in Fig.\ref{fig:five}f (see Methods for the CM definition) and is crucial for graphene's application in photovoltaics and photodetectors.

We emphasize that dynamical screening, when cured by cutting-off the singularity of the polarization function along the light cone with the parameter $\Lambda$, improves the agreement with experiments when compared to static screening only. Indeed, the static theory underestimates screening since it misses collinear scattering. This explains why $\Delta T/T$ calculated with static screening: i) increases too fast in the early stages, the electron dynamics being initially boosted by the poorly-screened Coulomb repulsion, and ii) slows down too much in the subsequent stages, when poorly-screened carriers begin to accumulate close the Dirac point.

In conclusion, we performed time-resolved spectroscopy on single-layer graphene with an unprecedented combination of temporal resolution and spectral tunability allowing us to track the early processes involved in electron thermalization. A microscopic theory based on the quantum Boltzmann equation and including collinear scattering and screening is capable of modeling the experimental data with no need of free parameters. In the region of parameter space explored in this experiment this ultrafast equilibration dynamics can only be explained by considering carrier multiplication and Auger recombination as fundamental mechanisms driven by electron-electron interactions.

We note that collinear Coulomb collisions in the intra-band scattering channel yield logarithmically-enhanced quasiparticle decay rates and transport coefficients (such as viscosities and conductivities)\cite{kashuba_prb_2008,fritz_prb_2008,schutt_prb_2011}. Angle-resolved ultrafast measurements of the hot-electron distribution may shed light on these important processes.

Ultrashort light pulses could be used to create a super-hot plasma of ultrarelativistic fermions (massless Dirac fermions) and bosons (e.g. phonons) in graphene or in other Dirac materials, thereby creating conditions analogue to those in early universe cosmogony, but within a small-scale, table-top experiment. Understanding the impact of collinear scattering on the ultrafast thermalization of massless Dirac fermions can thus be of pivotal importance to achieve a deeper understanding of high-temperature gauge theories\cite{arnold_jhep_2000,moore_jhep_2001}.

\begin{acknowledgements}
We thank Leonid Levitov, Allan MacDonald, and  Justin Song for very useful discussions. We acknowledge funding from MIUR ``FIRB - Futuro in Ricerca 2010" - Project PLASMOGRAPH (Grant No. RBFR10M5BT), the ERC grant NANOPOTS, EU grants RODIN and GENIUS, a Royal Society Wolfson Research Merit Award, EPSRC grants EP/GO30480/1 and EP/G042357/1, and the Cambridge Nokia Research Centre. Free Software (www.gnu.org, www.python.org) was used in this work.
\end{acknowledgements}

\section*{Methods}
\subsection*{Graphene growth and transfer}
SLG is first grown on copper foils (Cu) by Chemical Vapor Deposition (CVD)\cite{Li_s_2009,bae}. A $\sim 25~{\rm \mu m}$ thick Cu foil is loaded in a 4 inch quartz tube and heated to $1000$~$^{o}{\rm C}$ with an H$_{2}$ gas flow of 20 cubic centimeters per minute (sccm) at 200 mTorr. The Cu foils are annealed at $1000$~$^{o}{\rm C}$ for 30 mins. The annealing process not only reduces the oxidized foil surface, but also extends the graphene grain size. The precursor gas, a mixture of H$_{2}$ and CH$_4$ with flow rates of 20 and 40 sccm, is injected into the CVD chamber while maintaining the reactor pressure at 600 mTorr for 30 mins. The carbon atoms are then adsorbed onto the Cu surface, and nucleate SLG via grain propagation~\cite{Li_s_2009,bae}. Finally, the sample is cooled rapidly to room temperature under a hydrogen atmosphere at a pressure of 200 mTorr. The quality and number of layers of the grown samples are investigated by Raman spectroscopy\cite{RamanACF, cancado}. The Raman spectrum of graphene grown on Cu does not show any D peak, indicating the absence of structural defects. The 2D peak is a single sharp Lorentzian, which is the signature of SLG. We then transfer a $10 \times 10~{\rm mm}^2$ region of SLG onto quartz substrates ($100~{\rm \mu m}$ thick) as follows. Poly(methyl methacrylate) (PMMA) is spin-coated on the one side of graphene samples. The graphene films formed on the other side of Cu foil, where PMMA is not coated, is removed by using oxygen plasma at a pressure of 20 mTorr and a power of 10 W for 30s. Cu is then dissolved in a 0.2 M aqueous solution of ammonium persulphate ((NH$_4$)$_2$S$_2$O$_8$). The PMMA/graphene/Cu foil is then left floating until all Cu is dissolved. The remaining PMMA/graphene film is cleaned by deionized water to remove residual salt. Finally, the floating PMMA/graphene layer is picked up using the target quartz substrate and left to dry under ambient conditions. After drying, the sample is heated to $180~^{o}$C for $20~{\rm min}$ to flatten out any wrinkles~\cite{Pirkle_apl_2011}. The PMMA is then dissolved in acetone, leaving the graphene adhered to the quartz substrate. A portion of the substrate is not covered with graphene, thus allowing the measurement of the nonlinear response of the substrate by a simple transverse translation of the sample. This contribution is measured to be negligible.

The transferred graphene is then inspected by optical microscopy, Raman spectroscopy and absorption microscopy. After transfer, the 2D peak is still a single sharp Lorentzian, indicating that SLG has been transferred. The absence of D peak proves that no structural defects are induced during the transfer process. Raman measurements over a large number of points indicate a$\sim$200meV p-doping\cite{Das_nn_2008,Casiraghi_apl_2007,Pisana_nm_2007}.
\subsection*{Pump-probe spectroscopy}
The transient absorption spectroscopy setup is driven by a regeneratively-amplified mode-locked Ti:Sapphire laser (Clark Instrumentation) that delivers $150~{\rm fs}$ pulses at $ 780~{\rm nm}$ with $500~{\rm mJ}$ energy at $1~{\rm kHz}$ repetition rate. The laser drives three optical parametric amplifiers (NOPAs), from which the visible pump pulses and the two near-IR probe pulses are generated. These are then compressed to the transform limit duration by means of custom made chirped mirrors (visible NOPA), a fused silica prism pair (IR NOPA 1) and an adaptive shaper based on a deformable mirror (IR NOPA 2). The pump and probe pulses are synchronized by a motorized translation stage and spatially overlapped on the sample in a slightly noncollinear geometry. After the sample, the probe beam is focused onto the entrance slit of a spectrometer equipped with a 1024 pixel linear Si photodiode array (Entwicklungsbuero Stresing). The IR NOPA 2 probe pulse is instead detected by an InGaAs CCD spectrometer (Bayspec Super Gamut). Both this devices allow a full $1$-${\rm kHz}$ readout of the spectra. By recording pump-on and pump-off probe spectra, we extract the differential transmission signal as a function of pump-probe delay ($t$) as $\Delta T/T(\lambda,t) = [T_{\rm on}(\lambda,t) - T_{\rm off}(\lambda,t)]/T_{\rm off}(\lambda,t)$. The system has a sensitivity better than $\Delta T/T=10^{-4}$. The pump intensity was reduced to avoid any sample saturation or high order non-linear effects ($I_{\rm pump}<10~{\rm J}~{\rm cm}^{-2}$). By moving from multichannel to single-wavelength detection, we were able to reduce the fluence by a factor $20$, and saw a substantially unchanged dynamics. The amplitude of the $\Delta T/T$ signal is lower than $0.007$ at the maximum of the PB signature and the signal from the substrate is negligible.
\subsection*{Quantum Boltzmann Equation}
The QBE for the electron distribution,
\begin{equation}\label{eq:EOMel}
\frac{d f_{\mu}(\varepsilon_{{\bm k}, \lambda})}{dt} = \left.  \frac{d f_{\mu}(\varepsilon_{{\bm k}, \lambda})}{dt} \right|_{\rm e-e} + \left.  \frac{d f_{\mu}(\varepsilon_{{\bm k}, \lambda})}{dt} \right|_{\rm e-ph}~,
\end{equation}
includes collisions integrals due to e-e and e-ph scattering. The magnitude of the e-ph couplings is discussed in Refs.\onlinecite{PLMFR2004,lazzeri_prb_2008,basko_prb_2009}. Here we use the values $\langle g^{2}_{\Gamma} \rangle = 0.0405~{\rm eV}^{2}$ for the phonons at ${\bf \Gamma}$ and $\langle g^{2}_{\rm K,L} \rangle = 0.00156~{\rm eV}^{2}$, $\langle g^{2}_{\rm K,T} \rangle = 0.2~{\rm eV}^{2}$ for the longitudinal and transverse phonons at ${\bf K}$, respectively.

The equation for the phonon distribution,
\begin{equation}\label{eq:EOMph}
\frac{d n_{\bm q}^{(\nu)}}{dt} = \left. \frac{d n_{\bm q}^{(\nu)}}{dt} \right|_{\rm e-ph} - \frac{\gamma_{\rm ph}}{\hbar}[n_{\bm q}^{(\nu)} - n_{\rm eq}^{(\nu)}]~,
\end{equation}
includes the collision term due to e-ph scattering and the linear relaxation with $\gamma_{\rm ph} / \hbar \simeq 0.26~{\rm ps}^{-1}$~\cite{sun_prb_2012}. The equilibrium phonon distribution function consists of the Bose-Einstein thermal factor $n_{\rm eq}^{(\nu)} = \{\exp{[\hbar\omega^{(\nu)}_{\bm q}/(k_{\rm B}T)]}-1\}^{-1}$ evaluated at the $\nu$-th phonon branch
$\omega^{(\nu)}_{\bm q}$, assumed dispersionless in the present treatment. This approximation is well justified since $\omega^{(\nu)}_{\bm q}$ changes slowly with respect to the electron dispersion.

The Coulomb kernel in Eq.~(\ref{eq:circularQBE}) reads:
\ber\label{eq:coulombkernelexplicit}
{\cal C}_{\mu}(\varepsilon_{1}, \varepsilon_{2} , \varepsilon_{3}) &=& \frac{2\pi}{\hbar} \frac{1}{S^2} \sum_{{\bm Q}, {\bm k}_{3}}
|V^{(\mu)}_{1234}|^2\nonumber\\
&\times&\delta(|E - \varepsilon_{1}| - \hbar v |{\bm Q} - {\bm k}_{1}|)\nonumber\\
&\times& \delta(|\varepsilon_{3}| - \hbar v k_{3})\nonumber\\
&\times&\delta(|E - \varepsilon_{3}| - \hbar v |{\bm Q} - {\bm k}_{3}| + \eta)~,
\eer
where $E \equiv \varepsilon_{1} + \varepsilon_{2}$, ${\bm k}_{2} \equiv {\bm Q} - {\bm k}_{1}$, ${\bm k}_{4} \equiv {\bm Q} - {\bm k}_{3}$, and $S$ is the sample area. The polar angle of ${\bm k}_1$ does not matter, while the modulus of ${\bm k}_1$ is equal to $\varepsilon_1/(\hbar v)$. The Dirac delta functions follow from the conservation of total energy $E$ and momentum ${\bm Q}$.
In Eq.~(\ref{eq:coulombkernelexplicit}) we introduced the infinitesimal $\eta$ in the argument of the third delta to relax energy conservation, which is restored by taking the limit $\eta \to 0$. As shown in Fig.~\ref{fig:four}c), when $\eta = 0$ the summand in Eq.~(\ref{eq:coulombkernelexplicit}) vanishes for Auger processes. In this case, it is important to {\it first} perform the summations over ${\bm Q}$ and ${\bm k}_3$, and {\it then} take the limit $\eta \to 0$.

The squared matrix element $|V^{(\mu)}_{1234}|^2$ (where the integers $1 \dots 4$ indicate the dependence on $\lambda_{i}$ and ${\bm k}_{i}$ for $i = 1 \dots 4$) includes a summation over spin degrees-of-freedom and direct and exchange~\cite{Giuliani_and_Vignale} contributions to e-e scattering. It reads $|V^{(\mu)}_{1234}|^2 = |{\cal U}^{(\mu)}_{1234} - {\cal U}^{(\mu)}_{1243}|^{2}/2 + |{\cal U}^{(\mu)}_{1234}|^{2}$, where
\be\label{eq:Coulombmatrixelement}
{\cal U}^{(\mu)}_{1234} = W(|{\bm k}_{1} - {\bm k}_{3}|, \omega;t) F^{(\mu)}_{\lambda_{1},\lambda_{3}}(\theta_{{\bm k}_{3}} - \theta_{{\bm k}_{1}}) F^{(\mu)}_{\lambda_{2},\lambda_{4}}(\theta_{{\bm k}_{4}} - \theta_{{\bm k}_{2}})
\ee
is the matrix element of the Coulomb interaction in the eigenstate representation of the MDF Hamiltonian\cite{castroneto_rmp_2009}, with $F^{(\mu)}_{\lambda,\lambda'}(\theta) = [1 + \lambda \lambda' \exp{(i \mu \theta)}]/2$ the so-called ``chirality factor"\cite{castroneto_rmp_2009} and $\omega=(\varepsilon_{1}-\varepsilon_{3})/\hbar$. Note that Coulomb scattering occurs only {\it within} a valley $\mu$.

The contribution due to Auger processes to the Coulomb kernel can be calculated analytically:
\be\label{Augercontribution}
\left.{\cal C}_{\mu}(\varepsilon_{1},\varepsilon_{2},\varepsilon_{3})\right|_{\rm Auger} = \frac{1}{8\pi^2 \hbar^5 v^4} \sqrt{ \left | \frac{\varepsilon_{2}\varepsilon_{3}\varepsilon_{4}}{\varepsilon_{1}} \right |} |V^{(\mu)}_{1234}|^{2}~.
\ee
The strength of e-e interactions in graphene is parametrised by the dimensionless fine structure constant $\alpha_{\rm ee} \equiv e^{2} / (\hbar v \bar{\epsilon})$. We use $\alpha_{\rm ee} = 0.9$, as appropriate\cite{kotov_rmp_2012} for graphene with one side exposed to air and the other to ${\rm SiO}_2$.

The differential transmission is calculated as a function of wavelength $\lambda = 2\pi c/\omega$ from the electron distribution via the following relation~\cite{BKWMM2011}:
\ber
\frac{\Delta T}{T}(\lambda,t) &=& \pi \alpha [f_{\mu}(\hbar\omega / 2) - n_{\rm F}(\hbar\omega/ 2) \nonumber\\
&-&  f_{\mu}(-\hbar\omega / 2) + n_{\rm F}(-\hbar\omega / 2)]~,
\eer
where $\alpha = e^2/(\hbar c) \simeq 1/137$ is the fine-structure constant and $n_{\rm F}(E)$ is the Fermi-Dirac distribution. Here $\mu=\pm 1$ is not summed over and can be chosen at will, since the electron distribution is identical for the two valleys.

The CM is calculated by employing the following equation:
\begin{equation}
{\rm CM} = \frac{n_{\rm c}(t) - n_{\rm c}(-\infty)}{n_{\rm c}(0) - n_{\rm c}(-\infty)}~,
\end{equation}
where $n_{\rm c}(t) = \sum_{\mu} \int_{0}^{\infty} d\varepsilon f_{\mu}(\varepsilon) \nu(\varepsilon)$ is the electron density in conduction band at time $t$, $\nu(\varepsilon) = 2 \varepsilon {\cal A}_{0} / [2 \pi (\hbar v)^{2}]$ being the MDF density-of-states, and ${\cal A}_{0} \simeq 0.052~{\rm nm}^{2}$ the area of the elementary cell.

The numerical solution of the QBE is performed with a fourth-order Runge-Kutta method. The electron energies are discretized on a mesh with a $25~{\rm meV}$ step. The screening function and the Coulomb kernel are updated in time at multiples of the integration step, depending on the speed of the relaxation, with more frequent updates ({\it e.g.} every $2~{\rm fs}$) at the beginning of the time evolution.

\end{document}